\documentclass[reprint,showpacs,amsmath,amssymb,aps]{revtex4-1}
\pdfoutput=1

\usepackage{graphicx}
\usepackage{dcolumn}
\usepackage{bm}

\begin{document}

\title{$2K(2\nu)$-Capture in $^{124}$Xe: \\
Results of Data Processing
for an Exposure of 37.7 kg$\times$day
}

\author{
Yu.M.~Gavriljuk$^{a}$,
A.M.~Gangapshev$^{a}$,
V.V.~Kazalov$^{a}$,
V.V.~Kuzminov$^{a}$,
S.I.~Panasenko$^{b}$,
S.S.~Ratkevich$^{b}$ and
D.A.~Tekueva$^{a}$%
}
\affiliation{\small $^a$ Institute for Nuclear Research, RAS, Russia \\
\small $^b$ V.N.~Karazin Kharkiv National University, Ukraine}


\begin{abstract}
The results of the experimental search for two-neutrino $2K$-capture in $^{124}$Xe with a large copper proportional counter obtained by processing the data for an exposure of 37.7 kg$\times$day are presented. The experimental setup is located at the Underground Low-Background Laboratory of the Baksan Neutrino Observatory at a depth of 4900 m w.e. The combination of methods of selection of useful signals with a unique set of characteristics and the event topology taken into account allowed us to suppress the background in the energy region of interest. A new half-life limit for $2K(2\nu)$-capture in $^{124}$Xe was determined: T$_{1/2}\geq7.7\cdot10^{21}$ yrs (90\%
 C.L.).
\end{abstract}


\maketitle

\section{\label{Intr}Introduction}
The simultaneous capture of two electrons from an
atomic shell is a rare nuclear decay that may occur in
isotopes located at the right side of the mass parabola
for even-even isobars. In the case of $^{124}$Xe, the total
transition energy was measured with the Penning-trap
mass spectrometer SHIPTRAP to be 2856.73(12) keV
with an uncertainty of several hundred electronvolts
\cite{r1}. The capture of two bound electrons by a nucleus
with the emission of two neutrinos ($2\nu$ECEC) is
allowed in the Standard Model, while the lepton-nonconserving
neutrinoless transition ($0\nu$ECEC) is outside
the framework of the Standard Model and attracts
particular attention. However, although the transition
energy is rather high, even the search for $2\nu$ECEC
decays is a challenging experimental task, since only
the products of relaxation of the excited atomic shell of
a daughter atom are detectable.

The primary contribution (76.7\%
) to the ECEC process in $^{124}$Xe is produced by the capture of two electrons
from the $K$-shell \cite{r2}. The result of the
$^{124}$Xe$(2e_K,2\nu _e)$ reaction is a neutral $^{124}$Te atom with a
``lifted'' shell, which leaves both $K$-shell vacancies
exposed. The residual excitation of the atomic shell in
daughter isotope $^{124}$Te$^{**}$ relaxes via the emission of
Auger electrons $(e _A, e _A)$, a single characteristic quantum
and an Auger electron $(K, e_A)$, or two characteristic
quanta and low-energy Auger electrons $(K, K, e_A)$.
In actual experiments, the almost simultaneous emission
of two characteristic fluorescence quanta produced in the filling of two vacancies offers considerable advantages in terms of detection of such events.

The fluorescence yield in the process of filling of a
single $K$-shell vacancy by electrons from the upper
shells is $\omega _K$(Te) = 0.857. If two vacancies are present,
the probability of a radiative transition is, in a first
approximation, simply doubled. The probability of the
Auger transition also increases by a factor of two.
Thus, the emission of two characteristic photons produces
a considerable contribution to the atomic-shell
deexcitation in $^{124}$Te$^{**}$. This allows us to use the topological
method of identification of a complex event in
a gas chamber, which has already been applied in the
search for $2K$-capture in $^{78}$Kr \cite{r3}.

If we represent the process as two consecutive fillings,
the first photon is emitted in the transition of an
electron from the upper shells (e.g., $L$-shell) to one of
the two $K$-vacancies. The second photon is emitted in
the transition to the remaining $K$-vacancy (e.g., the
transition of an electron from the same $L$-shell with
one vacancy). The first photon is often called the
hypersatellite $(H)$, and the second photon is the satellite
$(S)$ \cite{r4}. Their energies are roughly equal to the
energies of common $K_{\alpha,\beta}$ - fluorescence quanta. More
accurate values were calculated in \cite{r5}, where the
RAINE package and the Dirac-Fock method were
used. The energy of the first fluorescence quantum
is ~680 eV higher than the $K_\alpha$ energy. The difference
for the second quantum is not that significant:
it is $\sim165$ eV harder than a common fluorescence
quantum. The yield of fluorescence in the form of
emission of two $K^H_\alpha$ and $K^S_\alpha$ photons was also determined
in \cite{r5}.

\section{EXPERIMENTAL PROCEDURE}

The experimental setup is exactly similar to the
``$2K$-Capture'' setup used to study the
$^{78}$Kr($2e_K, 2\nu _e$)$^{78}$Se reaction. It features a copper proportional
counter (CPC) with low-background shielding
(18 cm of copper, 15 cm of lead, and 8 cm of
borated polyethylene). The CPC is a 710-mm-long
cylinder with a working length of 595 mm and an inner
diameter of 137 mm. The anode wire made of goldplated
tungsten with a diameter of 10 $\mu$m is stretched
along the cylinder axis. A voltage of 2800 V is applied
to the anode. The total counter volume is 10.37 L, and
the working volume is 8.77 L. Xenon enriched to 21\%
in $^{124}$Xe was used as the working gas ($P_{gas}=4.8$ Bar). The
experimental setup and the method for signal
extraction were described in detail in \cite{r6}.

Our experiment requires stability of the spectrometric
detector parameters. The knowledge of time
variation (in long-term measurements) of such
parameters as the gas gain, the time of electron drift
from primary ionization to the region of proportional
multiplication, the ratio of amplitudes of the primary
and the secondary pulses, and the energy resolution in
the region of search for the effect of interest becomes
especially important. Although the counter is sealed,
microscopic amounts of quenching and accelerating
impurities enter the working gas during long-term
operation, which results in gradual deterioration of
spectrometric properties of the working gas. The cause
of this is the slow sublimation of polyatomic gas molecules
originating from the surfaces of the case and
sealers; in addition, oxygen diffuses gradually from the
flange sealing rings of the CPC and acts as a ``cooling''
additive. In order to preserve the operating detector
characteristics, the working gas was cleaned of electronegative
impurities (O$_2$ and N$_2$) every two months by
flushing the gas through a Ti reactor at $800^\circ$Ñ.

Calibration measurements with an external radioactive
source were conducted every two weeks to monitor
the spectrometric CPC parameters. In these measurements,
the CPC was irradiated through the wall
with $\gamma$-quanta from $^{109}$Cd ($E_\gamma$=88 keV) that had passed
through the calibrating aperture at mid-length of the
detector. The counting rate was $\sim3.3$ s$^{-1}$.

\section{DATA ANALYSIS}

The presence of pulses from secondary photoelectrons
(afterpulses) knocked out of the copper CPC
wall by ultraviolet photons, which were produced in
the region of avalanche initiation in the process of
amplification of primary-ionization electrons, provides
an opportunity to introduce the geometric features
of the working detector region into analysis. For
example, we use the dependence of the ratio of amplitudes
of the first afterpulse and the primary pulse on
the distance along the anode wire from the site of
development of the primary pulse to the center of
the wire to select near-edge events. This distance sets
the solid angle of coverage of the internal surface of the
copper cathode cylinder, which is roughly equal to
$3.9\pi$ at the midpoint of the anode wire and to $2\pi$ at the
edges of the working region of the wire. The relative
number of photoelectrons knocked out of copper by
photons produced in the gas discharge in primary ionization
amplification varies with this solid angle.
The analytical dependence of the ratio of amplitudes
of the primary and the secondary pulses allows one to
correct the CPC response and exclude near-edge
events from further analysis.

The pulse-afterpulse delay is defined by the total
drift time of electron from the cathode to the anode. It
sets the duration of the time interval that covers completely
any single event regardless of the distribution of
primary ionization throughout the CPC volume. The
sought-for current pulses of three-point events produced
by $2K(2\nu)$-capture in the CPC are sums of different
combinations of three bell-shaped components
smeared due to diffusion.

In order to select these useful events from the entire
set of background data, one should correct the detector
response function for the time of data acquisition
between gas cleaning procedures (run). With this in
view, we have determined the total energy release, the
ratio of amplitudes of the afterpulse and the primary
pulse $(\lambda_ñ)$, the delay between the primary and the secondary
pulses $(\bigtriangleup \tau)$, and the energy resolution for onepoint
events using the calibration data obtained with
$^{109}$Cd. Figures $1a$ and $1b$ show the time variation of
parameters $(\lambda_ñ)$ and $(\bigtriangleup \tau)$ determined in the interval of energy
from 80 to 96 keV for seven runs.
\begin{figure}
\includegraphics[width=2.8in,angle=0.]{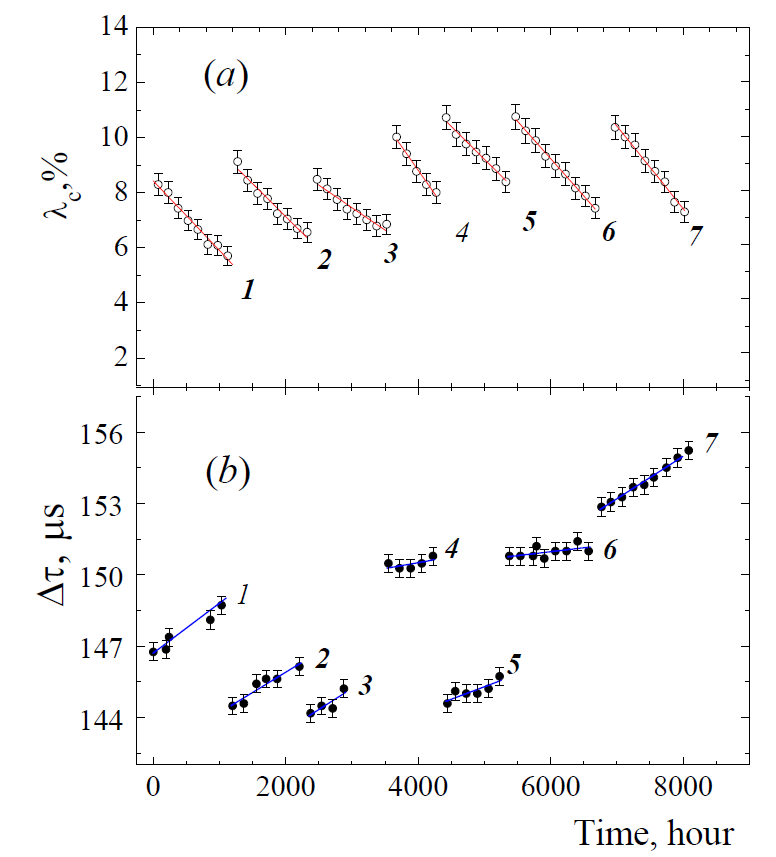}%
\caption{\label{f1}
Time variation of detector parameters between the
gas cleaning procedures for seven runs: (\emph{a}) ratio of the
amplitude of photoelectrons knocked out of the CPC case
by photons to the amplitude of complete absorption of a
photon with an energy of 88 keV ($^{109}$Cd); (\emph{b}) deterioration $\Delta \tau$ of the electron drift time from the cathode to the
anode.
}
\end{figure}

Multipoint events from the interaction of the working
gas with 88-keV photons were analyzed in order to
perform a more precise energy calibration of components
of a complex event. The data accumulated in
calibration measurements were processed, and the
detector responses were divided into groups with different
numbers of constituent partial pulses.

\begin{figure}[pt]
\includegraphics*[width=3.5in,angle=0.]{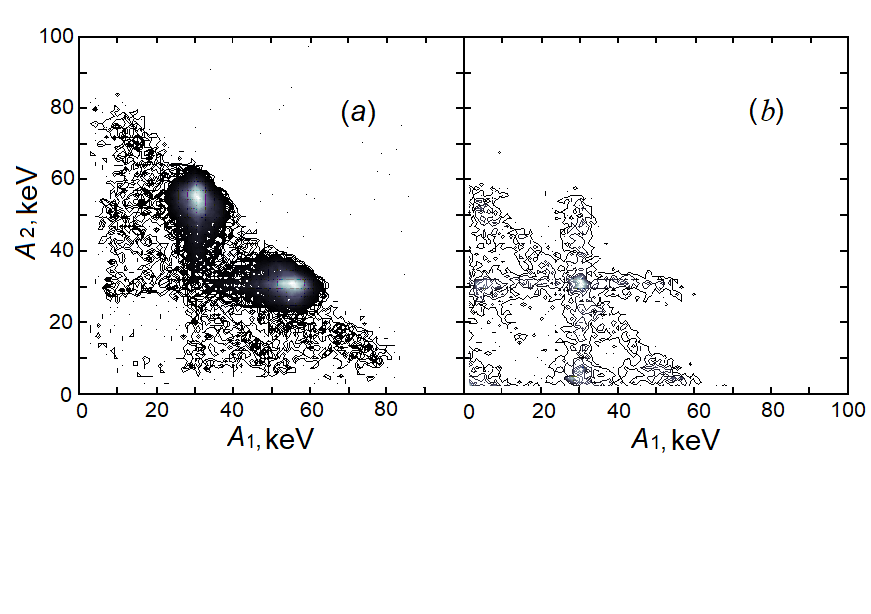}%
\caption{\label{f2} Two-dimensional amplitude distributions of the energy contributions of separate point ionization clusters from an external source of 88-keV photons ($^{109}$Cd): (\emph{a}) two-point events; (\emph{b}) three-point events after convolution with the amplitude of the third subpulse.}
\end{figure}
The group with a partial pulse contains those events where
the energy release from the absorption of an ionizing
particle is localized in a single point-like region of the
working detector volume. Such events are produced in
the absorption of $\alpha$-particles, electrons (without bremsstrahlung
photons), and, in part, X-ray or gamma
quanta. The absorption of quanta produces one-point
events in the following two cases: (1) Compton scattering
with the Compton photon escaping from the working
detector volume; (2) photoeffect accompanied
only by the emission of charged particles (photoelectron
+ Auger electrons).

Two-point events are produced
in the absorption of an electron with the emission
of a bremsstrahlung photon; in Compton scattering
of an X-ray or gamma quantum; or as a result of
the photoelectric effect accompanied by the emission
of a characteristic photon. It is required that the secondary
quanta be absorbed within the working detector
volume and form the second point-like ionization
region at a certain distance (in projection onto the
counter radius) from the first one. Two partial pulses
are produced for each ionization region as a result of
gas multiplication of primary-ionization electrons.
The delay between them is the difference between the
drift times of electrons through the distance from the
ionization region to the region of gas multiplication.
In xenon, the electron drift time from the cathode to
the anode, which defines the maximum time interval
of collection of primary ionization, is $\sim 147$ $\mu$s. If a photon
escapes from the counter, a one-point event is produced.
This event contributes to the escape peak of the
source line in the calibration spectrum.

Three-point
events may be observed in the case of complete
absorption (within the working detector volume) of
secondary particles produced (1) in Compton scattering
of an external quantum by an electron with the
absorption of the secondary quantum via photoeffect
accompanied by the production of a characteristic
$K$-quantum or (2) in $K$-shell ionization of the second
atom of the working medium by a $K$-photoelectron
from the primary photoeffect and filling of both
$K$-vacancies with the emission of characteristic photons.
In the case of double photoionization, threepoint
events are produced as a result of absorption of
two characteristic photons, which were produced in
the filling of a $2K$-shell vacancy in an atom of the studied
gas, and a group (photoelectron + Auger electrons)
originating from the same atom. All events with more
than three partial pulses are classified as multipoint
events. They may be produced, for example, as a result
of Compton scattering of $\gamma$-quanta by a $K$-electron
with subsequent absorption of the scattered photon via
$K$-shell photoeffect and filling two $K$-vacancies with
the emission of characteristic $K$-photons. Events may
migrate from one group to another due to the probable
overlap of partial pulses or the chance emergence of an
additional noise pulse.

The efficiency of selection of
one- and two-point events is determined by comparing
the result with calculated data obtained based on
the reference values of the fluorescence yield and the coefficient of absorption of intrinsic characteristic
radiation in the studied gas.

Figure $2a$ shows the intensity of two-point events as
a function of amplitudes of the first $(A1)$ and the second
$(A2)$ subpulses. The intensity maxima are
observed for a certain combination of subpulse amplitudes
($K_\alpha$(Xe)=29.8 and $88 - K_\alpha$(Xe)=58.2 keV).
Vertical and horizontal bands represent the specific
features of events with incomplete absorption of
88 keV of energy and the spectrum shape of input radiation.
Two diagonal bands are also visible. One of
them includes the events with complete absorption of
88 keV via Compton scattering and via photoeffect
with the emission of $e_{A(C)}$ and the production of
bremsstrahlung photons. The second band corresponds
to an energy of 58.2 keV and includes the
events from the escape peak of 29.8 keV from Compton
scattering of 88-keV photons by external electrons.

Figure $2b$ shows the two-dimensional distribution of
intensity of three-point events after folding over the
amplitude of the third subpulse $(A3)$. This distribution
reveals a peak at a combination of amplitudes $A1 = A2 =
29.8$ keV; vertical and horizontal bands at $A1 (A2) =
29.8$ keV formed by subpulses $A2 (A1)$ corresponding to
the energy depositions from Compton scattering of
88-keV quanta by external electrons; and a diagonal
band corresponding to $(A1 + A2) = 58.2$ keV at $A3 =
29.8$ keV.

The energy scale for amplitudes was calibrated
based on three points: 29.8, 58.2, and 88 keV. Only the
one-point pulse components were taken into account.
This revealed a certain nonlinearity of the amplitude
scale, which is associated with the fact that the voltage
operating mode of the counter corresponded to the
start of the limited-proportionality region, since
parameter $\lambda$ was required to have a sufficient value in
the interval from 20 to 80 keV for the purpose of selection
of multipoint events. The comparison of the
results of modeling of CPC processes with the distribution
of events from the external calibration $^{109}$Cd
source provides an opportunity to determine the coefficients
of selection of useful events.

\section{MEASUREMENT RESULTS}

\begin{figure}[pt]
\includegraphics*[width=2.9in,angle=0.]{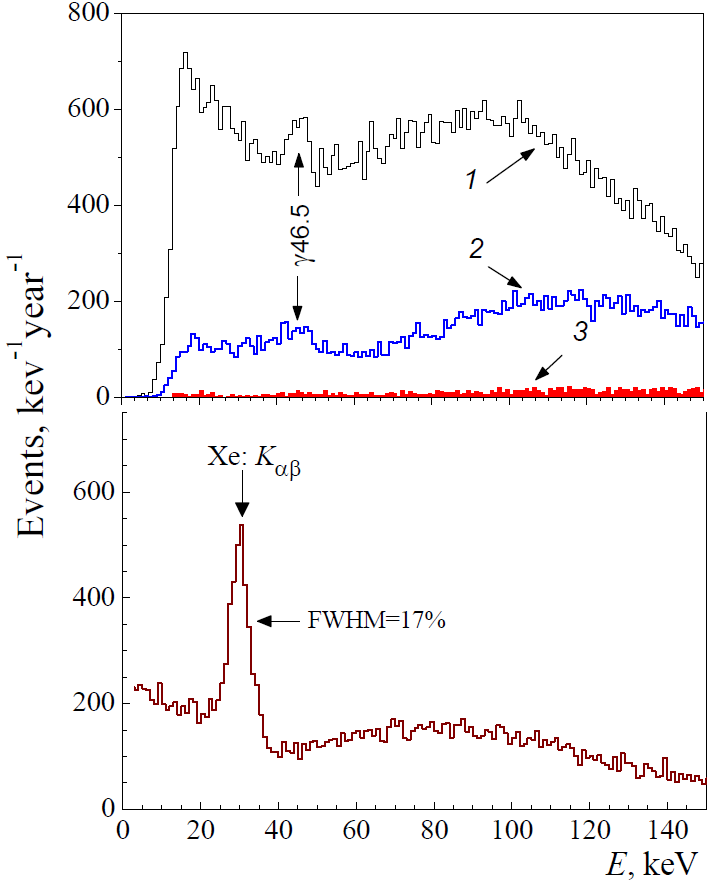}%
\caption{\label{f3} ($a$) Amplitude background spectra plotted for pulses with a front edge of 3.2-12 $\mu$s. Curves 1-3 correspond to the events with one, two, and three point clusters of ionization.
($b$) Amplitude spectrum of two-point events after
convolution with the amplitude of one of the two subpulses.}
\end{figure}
Figure $3a$ shows the resulting background spectra
of the CPC filled with the studied xenon gas normalized
to one year of measurements. Curves 1-3 correspond
to the events with one, two, and three point
clusters of ionization. The spectra were plotted based
on amplitudes of pulses with a front edge of 3.2-12 $\mu$s
with the CPC response function corrected for time in
a single run.

The 46.5-keV line corresponding to the
background $^{210}$Pb source is seen clearly in the amplitude
spectra of one- and two-point events. The $^{210}$Pb
isotope may be produced directly in copper in the
decay of radioactive microimpurities (volume source)
or be accumulated on the case surface (surface source)
in the process of detector fabrication as a result of
deposition of daughter products of decay of $^{222}$Rn
present in atmospheric air. Gamma radiation from
both sources enters the working counter volume. The
peak is formed primarily by one-point (xenon photoeffect
with deexcitation by Auger electrons) and
two-point events (photoeffect with the emission of
characteristic xenon radiation). A small fraction of
events with the absorption of the primary or characteristic
photon via scattering by external electrons with
subsequent absorption of the secondary quantum are
three-point ones.

The relatively large cross section of interaction of
background photons via photoeffect at the $K$-shell of
xenon yields another intense peak in the spectrum
after folding of two-point events. This background $K$-line
serves as an additional reference for amplitude
correction and a more precise determination of the
energy resolution in the entire data set. This is illustrated
by Fig. ~$3b$, where the integral peak of $K$-lines
$\{K_{\alpha 1} = 29.779$ (100\%),
$K_{\alpha 2} = 29.458$ (54\%),
$K_{\beta 3} = 33.562$ (9\%),
$K_{\beta 1} = 33.624$ (18\%),
$K_{\beta 2} = 34.415$ (5\%)$\}$
after Xe photoeffect with a resolution of 17\%
is seen clearly.

\begin{figure}[pt]
\includegraphics*[width=2.9in,angle=0.]{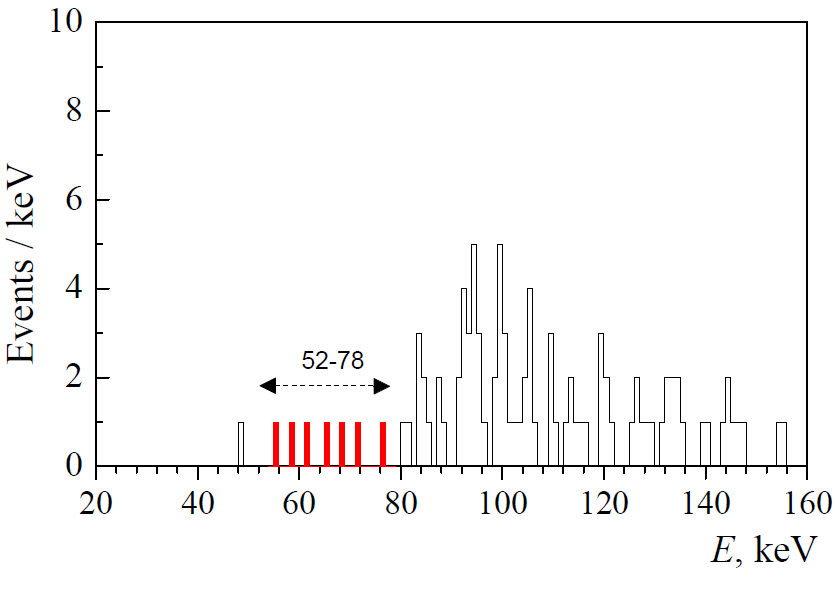}%
\caption{\label{f4} Resulting spectrum of three-point events with specific features obtained in 643 days of live measurements.}
\end{figure}
Figure 4 presents the resulting amplitude spectrum
of the total energy deposition in three-point events
selected from the entire data set formed in 643 days of
live measurements. These events were selected in
accordance with the following three criteria: (1)
energy deposition in one of the three ionization charge
clusters has an amplitude of 5-13 keV; (2) one of the
two remaining clusters has an amplitude of 21-35 keV;
(3) the third point charge cluster has an amplitude
above 13 keV. The second criterion was specified by
the energy intervals for $ K^{H(S)}_\alpha \pm 1.64 \sigma$, where $\sigma \sim 5$ keV
and $\{K^{H}_{\alpha 1} = 28.151$, $K ^{H}_{\alpha 2} =  27.858$,
$K^{S}_{\alpha 1} = 27.636$, $ K ^{S}_{\alpha 2} = 27.362$ keV \cite{r5}$\}$.

According to \cite{r1}, the total energy of the excited
atomic shell of $^{124}$Te$^{**}$ with two $K$-vacancies is
$Q_{KK}=64.457$ keV. A close value of $Q_{KK}=64.390$ keV was
obtained in \cite{r5} in calculations performed using the
multiconfiguration Dirac--Fock method. Using the
energy resolution parameters of the detector, we
define the region of interest as the interval of total
energy depositions ranging from 52 to 78 keV. A total
of seven events from the entire data set fell within this
region. The number of events is small at present, and
it is impossible to determine the signal-to-background
ratio correctly. Since these events may be either background
or useful ones, we identify $n_{exp}=7^{+5.5}_{-1.4}$ events
(90\%
CL) in accordance with the procedure outlined
in \cite{r7} within the energy range of our search for
$2K(2\nu)$-capture. Taking the upper limit on the number
of expected events $n_{exp} = 12.53$, we then obtain the
lower limit for the half-life of $^{124}$Xe:
\begin{equation*}
{\texttt{lim }} T_{1/2}^{ 2K(2\nu)}  = {\frac{ln 2 N_a \omega^{2K} \eta t_e} {{\texttt{lim }} n_{exp}}} \geq 7.7 \times 10^{21} \texttt{ yr},
\end{equation*}
where $N_a = 2.85 \times 10^{23}$ is the total number of $^{124}$Xe
atoms; $\omega^{2K} = \omega^{H} \times \omega^{S} = 0.772$ is the fluorescence yield
in the case of simultaneous emission of hypersatellite
and satellite photons; $\eta = \varepsilon p \varepsilon_3 \alpha _k k_\lambda$ is the overall efficiency of selection of three-point events ($\varepsilon_p = 0.809$ is
the probability of simultaneous absorption of two
$K$-photons, $\epsilon_3 = 0.51 \pm 0.05$ is the coefficient of identification
of three-point events, $\alpha _k = 0.985 \pm 0.005$ is
the coefficient of detection of $2K$-photons and Auger
electrons as three-point events, and $k_\lambda = 0.89 \pm 0.01$ is
the coefficient that takes the selection with respect to
parameter $\lambda$ into account); $t_e = 1.76$ years is the exposure
time.

\section{CONCLUSION}

A total of seven events were found in the energy
region of the expected effect in the spectrum of three-point
useful events selected from the data set corresponding
to an exposure of 37.7 kg$\times$day of the CPC
filled with xenon enriched to 21\%
in $^{124}$Xe at pressure
$P=4.8$ Bar. These events were used to set a new limit
on the half-life of $^{124}$Xe with respect to two-neutrino
$2K$-capture. The experiment was conducted at the
Baksan Neutrino Observatory. The ``$2K$-Capture''
experimental setup is installed at the Underground
Low-Background Laboratory at a depth of
4900 m w.e., where the flux of cosmic ray muons is
$(3.03 \pm 0.10) \times 10^{–9}$ cm$^{–2}$ s$^{–1}$ \cite{r8}.

\begin{center}
    {\textbf{ACKNOWLEDGMENTS}}
\end{center}

This study was supported by the Fundamental
Research Program ``Fundamental Properties of Matter
and Astrophysics'' of the Presidium of the Russian
Academy of Sciences and by the Russian Foundation
for Basic Research, project no. Mol-a 16-32-00018
(2016).


\begin{references}

\bibitem{r1} D.A.~Nesterenko, K.~Blaum, M.~Block, C.~Droese,
S.~Eliseev, F.~Herfurth, E.~Ramirez, N.~Minaya, Yu.~Novikov, L.~Schweikhard, V.M.~Shabaev,
M.V.~Smirnov, I.I.~Tupitsyn, K.~Zuber, and
N.A.~Zubova, Phys. Rev. C \textbf{86}, 044313 (2012).

\bibitem{r2} M.~Doi and T.~Kotani, Prog. Theor. Phys. 87, 1207
(1992).

\bibitem{r3} Yu.M.~Gavrilyuk, A.M.~Gangapshev, Dj.A.~Zhantudueva,
V.V.~Kazalov, V.V.~Kuz'minov, S.I.~Panasenko,
S.S.~Ratkevich, K.V.~Efendiev, and S.P.~Yakimenko,
Phys. At. Nucl. \textbf{76}, 1063 (2013).

\bibitem{r4} M.S.~Freedman, Annu. Rev. Nucl. Sci. \textbf{24}, 209 (1974).

\bibitem{r5} S.S.~Ratkevich, A.M.~Gangapshev, Yu.M.~Gavrilyuk,
F.F.~Karpeshin, V.V.~Kazalov, V.V.~ Kuzminov,
S.I.~Panasenko, M.B.~Trzhaskovskaya, and S.P.~Yakimenko,
Phys. Rev. C \textbf{96}, 065502 (2017).

\bibitem{r6} Yu.M.~Gavrilyuk, A.M.~Gangapshev, V.V.~Kazalov,
V.V.~Kuzminov, S.I.~Panasenko, S.S.~Ratkevich, and
S.P.~Yakimenko, Instrum. Exp. Tech. \textbf{53}, 57 (2010).

\bibitem{r7} G.J.~Feldman and R.D.~Cousins, Phys. Rev. D \textbf{57},
3873 (1998).

\bibitem{r8} Yu.M.~Gavrilyuk, A.M.~Gangapshev, A.M.~Gezhaev,
V. V. Kazalov, V. V. Kuzminov, S. I. Panasenko,
S.S.~Ratkevich, A.A.~Smolnikov, and S.P.~Yakimenko,
Nucl. Instrum. Methods Phys. Res., Sect. A \textbf{729}, 576
(2013).

\end{references}
\end{document}